# Optical and structural properties of ultra-thin gold films

*Anna Kossoy\*, Virginia Merk, Denis Simakov, Kristjan Leosson*[†]

Science Institute, University of Iceland, IS107 Reykjavik, Iceland

\* Current address: Weizmann Institute of Science, Rehovot 7610001, Israel,
e-mail: anna.kossoy@gmail.com

[†] Current address: Innovation Center Iceland, Arleynir 2-8, IS112 Reykjavik, Iceland
e-mail: kleos@hi.is

*Stéphane Kéna-Cohen*[††] *and Stefan A. Maier*

Experimental Solid State Group, Blackett Laboratory, Imperial College London, London SW7 2AZ, United Kingdom

[††] Current address: Department of Engineering Physics, École Polytechnique de Montréal, Montréal, Quebec H3C 3A7, Canada

**Abstract**

Realizing laterally continuous ultra-thin gold films on transparent substrates is a challenge of significant technological importance. In the present work, formation of ultra-thin gold films on fused silica is studied, demonstrating how suppression of island formation and reduction of plasmonic absorption can be achieved by treating substrates with (3-mercaptopropyl) trimethoxysilane prior to deposition. Void-free films with deposition thickness as low as 5.4 nm were realized and remained structurally stable at room temperature. Based on detailed structural analysis of the films by specular and diffuse X-ray reflectivity measurements, it is shown that optical transmission properties of continuous ultra-thin films can be accounted for using the bulk dielectric function of gold. However, it is important to take into account the non-abrupt transition zone between the metal and the surrounding dielectrics, which extends through several lattice constants for the laterally continuous ultra-thin films (film thickness below 10 nm). This results in a significant reduction of optical transmission, as compared to the case of abrupt interfaces. These findings imply that the atomic-scale interface structure plays an important role when continuous ultra-thin films are considered, e.g., as semi-transparent electrical contacts, since optical transmission deviates significantly from the theoretical predictions for ideal films.

## 1. Introduction

Ever since thin-film deposition techniques with sub-micrometer thickness control were developed, researchers have explored the question of how and when material properties (electrical, thermal, optical, magnetic, mechanical, chemical) of thin solid films deviate from the bulk properties of the deposited material.[1] Such differences in observed parameters may arise, e.g., due to different size, shape and orientation of crystal grains, presence of interfaces, quantum confinement, built-in stress, and/or due to morphological effects related to the particular mode of film growth.

The frequency-dependent dielectric function of a material is a bulk property that, in many cases, adequately describes the dielectric response of thin films, even when the film thickness is far smaller than the wavelength of the oscillating electromagnetic field. Nanometer-scale surface morphology can, on the other hand, strongly modify the dielectric response. This is particularly important in the case of metallic films in the optical regime.[2] Furthermore, even in the case of structurally perfect films, the (local) dielectric function is a classical concept that, in the nanometer limit must be reevaluated to account for spreading of the electron wave functions and finite-size effects.[3,4] Such considerations have in recent years received strong focus within the field of plasmonics.[5,6]

Experimentally, it is difficult to separate structural and physical modifications of the dielectric response where real metal surfaces are involved. At the sub-nm length scale, theoretical calculations and numerical simulations based on the use of a nonlocal dielectric function have predicted limitations in field enhancement, tunneling and smoothing of atomic-scale surface roughness.[3,7,8] In the case of ultra-thin single-crystal gold films, density-functional theory calculations reveal bandstructure changes and thickness-dependent optical anisotropy.[4] Experimental verification of such phenomena is challenging, however, as it requires structural control down to atomic dimensions over comparatively long length scales.

Gold is a particularly important metal for research and technology, due to its chemical inertness, high conductivity, high work function, large atomic mass, and favorable optical properties for certain applications. Chemical inertness, however, also translates to high surface and bulk diffusivity and poor adhesion. The possibility of realizing ultra-thin gold films on glass substrates and understanding their optical behavior is of substantial technological interest, e.g., for transparent electrical contacts, optical metamaterials and nanoplasmonic devices. In these cases, it is necessary to minimize spurious optical absorption and (in some cases) low-frequency electrical resistivity. In this context, it is also important to be able to use conventional metal deposition methods to obtain high-quality films on common transparent substrate materials, such as glass. Recently, ultra-thin (<10 nm) high-quality gold films on silicon[9] and glass[10] surfaces activated with mercapto-silanes were reported. The idea of depositing Au on activated surfaces terminated with SH-groups stems from the research of adsorbing alkane-thiol monolayers on Au surfaces[11] and was explored, e.g., in 1991 by Goss et al.[12] for deposition of relatively thick films (70 nm). In 1994, Dunaway and McCarley reported scanning-force microscopy studies of ultra-thin (<10 nm) gold films on silicon and oxidized silicon using various silane monolayers.[13] Later, Hatton et al.[14]

reported optical transmission measurements on ultra-thin gold films deposited on silanized glass, but did not compare the results to a detailed structural analysis or theoretically expected transmission spectra. The method of surface activation with mercapto-silane is easier and less costly than some other fabrication procedures reported recently[15,16] and does not compromise the optical quality of ultra-thin films, as is the case when metallic adhesion/seeding layers (typically Ti, Cr, Ge or Ni) are used.[17] It should be noted that a substantial amount of research has also been devoted to ultra-thin silver films and their optical properties, where many of the same problems and applications are encountered.[18-22]

In the present paper, we investigate ultra-thin gold films, deposited onto fused silica activated with mercapto-silane, and study their optical transmission properties. The film structure was analyzed in detail using X-ray reflectivity measurements. We achieve laterally continuous gold films down to 5.4-nm thickness on activated substrates. We show that in order to achieve continuous films in this thickness range, high deposition rates (10 Å/s) or, alternatively, high-energy irradiation prior to or during deposition must be used. We show that optical transmission at low film thickness differs from that of an ideal film of bulk permittivity, even in the case of laterally continuous films. We propose a simple model of graded interfaces to account for these results. These results are of general importance for optical devices incorporating ultra-thin gold films as well as for experimentalists searching for signatures of non-local and finite-size modifications of the dielectric function or other nanometer-scale features in the optical response of gold nanostructures.

## 2. Sample fabrication and experimental methods

Gold was deposited onto clean fused silica substrates and onto fused silica activated with (3-mercaptopropyl)trimethoxysilane (Sigma-Aldrich). Substrates were cleaned in an ultrasonic bath in acetone for 10 min and in isopropanol for 5 min, dried with a nitrogen gun and subsequently treated in a plasma asher (PlasmaPreen) in a 1:1 $O_2$/Ar mixture for 45 s. Substrates were activated by direct evaporation of (3-mercaptopropyl)trimethoxysilane in atmospheric conditions under a glass cover for 1 week. Treatment for shorter time (4 days) resulted in less continuous gold films. In a vessel saturated with (3-mercaptopropyl)trimethoxysilane vapor under low vacuum, however, pre-treatment can be carried out within a few hours. Rinsing with toluene and isopropanol (5 min each in ultra-sonic bath) before or after Au evaporation did not affect film continuity, roughness or adhesion. In order to test the effects of high-energy irradiation, activated fused silica was in some cases exposed to UV radiation (up to 4.89 eV) from an Hg/Ar lamp at 0.15 W/cm$^2$ for 5 min prior to thermal deposition.

Metal deposition was performed by conventional thermal or electron-beam evaporation. An evaporation rate of about 1 Å/s and a base pressure of around 3 x 10$^{-6}$ mbar were used in most cases. The mass-equivalent thickness during deposition was recorded with a calibrated QCM (quartz crystal microbalance) detector. For each thickness, evaporation onto activated substrates and untreated reference substrates was done simultaneously to ensure identical deposition thickness.

Optical transmission was measured in the wavelength range from 400 nm to 1000 nm. All experimental and theoretical transmission curves shown below include a ≈4% reduction in overall transmission due to the incoherent reflection from the backside of the substrate. Optical transmission spectra were simulated using the transfer-matrix method. The permittivity values used in the calculation were obtained from thicker films deposited using the same equipment by simultaneously fitting variable-angle ellipsometry data (VASE, JA Woollam) and transmission data using a Kramers-Kronig-consistent generalized oscillator model. The resulting values agree closely with permittivity data recently reported by Olmon et al.[23] Angle- and polarization-dependent transmission measurements were conducted for continuous films.

Scanning electron microscope (SEM) images were taken with an in-lens secondary-electron detector on a Zeiss LEO SUPRA25 microscope. Transmission spectra of structured gold films were simulated with commercial finite-difference time-domain software (Lumerical FDTD Solutions). Transmission through a 300-nm square section of gold film was modeled, using a mesh size of 1 nm in the lateral direction and 0.5 nm along the propagation direction in the vicinity of the gold film. Image processing of SEM images to obtain binary masks for numerical simulation was made with an algorithm described in Ref. [24].

X-ray reflectivity (XRR) measurements were made with a parallel beam on a PANalytical MRD diffractometer with a Cu tube equipped with hybrid Ge[220] 2-bounce monochromator on the incident side (wavelength 0.154 nm). A 1/16° slit was mounted before the monochromator. A parallel plate collimator with an angle of acceptance 0.27° and receiving slit of 0.27° was mounted before the detector. Fitting of XRR spectra was done with X'Pert Reflectivity 1.1 PANalytical software which uses the Parratt formalism to model reflectivity spectra and the Nevot-Croce model to fit roughness.[25,26] Further details of the procedures employed in the fitting software can be found on the manufacturer's website.[27]

For Au films deposited on surface-activated substrates, rocking curves were measured at 1.4° for a reference sample without Au, at 2° and 3° for 3-nm Au; at 1.6°, 2.3°, 3.8° for 4.2-nm Au; at 1.6°, 2.3° and 3.8° for 5.4-nm Au; at 1.4°, 1.89° and 2.94° for 7.2-nm Au. Density, thickness and initial roughness values were obtained from fitting of specular reflectivity spectra. Roughness values were refined in fitting of rocking curves together with lateral correlation lengths, which were adjusted for each layer and Hurst parameter, which is specified for the whole sample. One set of those parameters describes all rocking curves for a particular sample. Rocking curves (RC) were fitted with the same software as specular reflectivity spectra, using fractal roughness modeled with distorted-wave Born approximation. Surface coverage of the gold film was determined as described in Ref. [24].

3. **Results and discussion**

Figure 1, panels a) and b), show results of transmission measurements for ultra-thin gold films deposited with an electron-beam evaporator onto activated and untreated substrates, respectively. Thickness values refer to the mass-equivalent-thickness as measured by the QCM, rather than the physical thickness of the resulting film as measured by XRR. It is immediately obvious that there is substantial difference in the optical properties of the resulting gold films. Gold films deposited on untreated substrates (Figure 1b) show a clear dip in transmission that red-shifts and widens with increasing deposition thickness. This is the standard signature of localized surface plasmon resonance absorption in discontinuous layers (island films) that are formed when gold is deposited on glass and many other types of substrates, as has been well known for decades.[28] This is corroborated by X-ray analysis, revealing surface coverage ranging between 35% and 75% for the deposited thicknesses. Gold films deposited on substrates activated with (3-mercaptopropyl)trimethoxysilane (Figure 1a), however, show a strikingly different behavior, with no indication of a transmission dip for deposition thickness of 5.4 nm and above. This is supported by scanning electron microscope investigations, confirming that a continuous featureless film has formed at this thickness (Figure 2). Importantly, however, the overall transmission across the measured wavelength range is significantly lower than expected for ideal flat films described by the bulk dielectric function of gold of the same mass-equivalent thickness (Figure 1c), as will be discussed in more detail below, while transmission through thicker films corresponds very closely to the theoretical prediction (Figure 3).

On one hand, we find that continuous ultra-thin films can be realized on activated surfaces by using high Au evaporation rates (e.g. 10 Å/s), using a thermal (resistive) evaporator. At such high rates, however, accuracy in controlling the deposition thickness of ultra-thin films is significantly reduced. On the other hand, using the e-beam deposition method can ensure high-quality gold films at much lower evaporation rates. In Figure 1a, we demonstrate the difference in transmission spectra of Au films with the same mass-equivalent thickness of 5.4 nm deposited by e-beam and thermal evaporation, both at approximately 1 Å/s, on substrates treated with (3-mercaptopropyl)trimethoxysilane. Thermally deposited Au on treated substrates exhibits a transmission spectrum similar to films deposited on untreated substrates (Figure 1b) for comparable evaporation rates. We attribute this difference to the exposure of the substrate to high-energy radiation (electromagnetic and/or secondary electron emission) during the e-beam process. This assertion is supported by comparison between 3.0-nm films deposited by thermal evaporation onto substrates with or without prior exposure to UV radiation from an Hg(Ar) lamp (mainly 253.65 nm) which resulted in a visible difference in color of the deposited films (Figure 4). Thiol-groups are well known to participate in photochemically triggered click-reactions with allyls,[29] while it is the double bond (-C=C-) which is believed to be activated by UV radiation. Our findings suggest that the bond between sulfur and proton is weakened by high-energy radiation leading to increased reaction with Au. The mechanism can be described as:

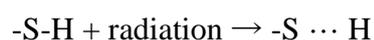

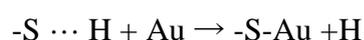

The reaction with the thiol-group reduces Au surface diffusion, producing nucleation sites leading to formation of a laterally continuous film at a critical thickness of no more than 5.4 nm. Similarly improved gold attachment has also been observed on certain types of optical polymers[30] but with higher critical thickness.

In order to extract more information about the film structure, we performed detailed XRR studies of the gold films deposited on activated substrates, including measurements of diffuse scattering around several diffraction angles.[31] Figure 5 demonstrates typical reflectivity spectra and rocking curves together with their fits, obtained for 3.0-nm and 7.2-nm-thick Au films (values refer to mass-equivalent thickness). Importantly, the measured XRR data could not be consistently fitted by assuming only a single Au layer (with adjustable density, thickness and roughness). Instead, we describe the gold film as consisting of two layers, which gives consistently good fits for all measured samples. We refer to these "surface" and "bulk" layers of the gold film in the fitting procedure as "top" and "bottom" gold layers, respectively.

Results of the fits for the measured films are summarized in Figure 6, and the layer model is illustrated in the inset. Parameters corresponding to the top and bottom gold layers are shown separately. The lateral correlation length was large (typically around 100 nm) compared to the total layer thickness in all cases and the fractal (Hurst) parameter was similar for all samples, in the range 0.40-0.55. The fitting also takes into account the possibility of an adsorbed water layer on top of the gold film, which for all cases was found to be ≈1.5 nm in thickness with a roughness of 0.2 nm. XRR-determined roughness of the substrate was $r_{sub}$=1.0±0.4 nm, while roughness of the top layer was $r_{top}$=0.6±0.1 nm. In all cases, the top Au layer was observed to have lower density than the bottom layer, reaching a maximum of ≈17 g/cm$^3$ for continuous 5.4 and 7.2 nm films. For 5.4 and 7.2 nm (and thicker) films, the density of the bottom layer matches that of bulk Au (19.3 g/cm$^3$). The thickness of the surface layer derived from the double-layer fit starts to drop once the film becomes continuous and for thicker films (>12 nm), XRR spectra can be accurately fitted by assuming only a single layer with a certain degree of surface roughness, typically around 1 nm.

As an independent check of consistency, we calculate the total mass-equivalent thickness determined from the XRR fitting procedure d=($d_{top}*\rho_{top}+d_{bot}*\rho_{bot}$)/$\rho_{bulk}$, as shown in Figure 6b. We find that this value is proportional to the deposition thickness and agrees with the independently calibrated QCM values to within one atomic layer of gold. This confirms that our XRR fitting procedure does not over- or underestimate the total amount of gold deposited, but gives a more detailed picture of the cross-sectional density variation through the gold layer than the simple model of a perfectly dense film with surface roughness which often is assumed in literature, where information about film structure is often derived solely from surface-probe measurements such as atomic force microscopy (AFM).

SEM investigation reveals that coalescence of nucleated gold islands occurs already at a deposition thickness below 3 nm (see also inset of Figure **7**a). The corresponding coalescence thickness for a clean $SiO_2$ surface is at least two times larger, when using the same deposition method.[30] On activated glass, the transition from coalescence to full continuity occurs in the thickness range between 3 nm and 5.4 nm (again, the continuity thickness is at least two times lower than on untreated $SiO_2$).

We now proceed to analyzing the measured optical transmission spectra shown in Figure 1a. As is already clear from comparison of Figures 1a and 1c, there is significant discrepancy between measured data and the theoretical transmission spectra for perfect films of the same mass-equivalent thickness. It is interesting to note, however, that the transmission spectra for the continuous ultra-thin films are qualitatively similar to the modeled result, except for a vertical offset which is remarkably constant across most of the measured wavelength range. For an ideal film, as film thickness increases, transmission drops faster at longer wavelengths than around the surface plasmon resonance frequency and, therefore, the observed optical (VIS-NIR) transmission cannot be accounted for by a single-layer model. Furthermore, the constant shift (also observed at frequencies above the surface plasmon resonance frequency) makes it unlikely that the difference can be attributed to absorption by localized surface plasmon resonances, related to morphological features. The observed differences do not match the change in permittivity that has been predicted by DFT in single-crystal (111) gold films of similar thickness[4] and we observed no sign of optical anisotropy in angle- and polarization-dependent transmission spectra of the laterally continuous films.

Considering the results of the XRR fitting procedure discussed above, we also compared the measured data to a two-layer model where the reduced gold density was included via the Bruggeman approximation[32]

$$f \frac{\varepsilon_a - \varepsilon_{\text{eff}}}{\varepsilon_{\text{eff}} - L(\varepsilon_a - \varepsilon_{\text{eff}})} + (1-f) \frac{\varepsilon_b - \varepsilon_{\text{eff}}}{\varepsilon_{\text{eff}} - L(\varepsilon_b - \varepsilon_{\text{eff}})} = 0, \qquad (1)$$

where the filling fraction $f$ corresponds to the gold ($\varepsilon_a$) density relative to the bulk value, in a surrounding dielectric ($\varepsilon_b$). Depolarization factors $L$ corresponding to 2D and 3D inclusions were tested. Also, we simulated by FDTD, the effect of a structured top layer on a fully dense bottom layer. These calculations, however, only resulted in minor deviations from the perfect-film model and are not shown here.

Considering the failure of the above approaches to reproduce the observed data, we introduce a new model that successfully accounts for the observed transmission spectra across the measured wavelength range. In this case, the film is modeled using its measured thickness and the bulk dielectric function of gold (for accuracy, we used the dielectric function measured on 30-nm thick films deposited using the same equipment, as described above). In the model, the roughness of the top layer and the substrate is represented as additional layers, described by the effective-medium dielectric function of gold ($f$=0.5) in air or glass, respectively, using Equation (1). For the thickness of these interface layers, values of $r_{\text{sub}}$ and $r_{\text{top}}$, respectively, derived from the XRR fit, were used. The inclusion of such interface layers in the model results in increased absorption and reflection by the gold film over a wide wavelength range and, as shown in Figure 7, correctly predicts the shape of the transmission spectra for continuous films (Figures 7c and 7d). Furthermore, this is supported by our previous studies which show that the agreement between measured data and ideal film behavior gradually increases as gold film thickness approaches 15-20 nm.[30] Finally, it agrees very well with the recent findings reported by Slovinsky et al.[33] where anomalously large propagation loss of long-range surface plasmon polaritons on ultra-thin gold films at 1550 nm wavelength was attributed to the presence of similarly graded interface layers.

Obviously, the simple transfer-matrix method does not correctly account for plasmonic absorption due to cracks through the gold film, which are present in the coalescence regime, in samples having a gold film thickness between 3 nm and 5.4 nm. In order to include the effect of the structure of non-continuous films on the transmission spectra, we performed FDTD simulation based on filtered SEM images, as described in the previous section, with examples shown as insets in Figures 7a and 7b. A similar approach was recently used to investigate SERS enhancement in island films[34] but the authors did not study the related effect on optical transmission through the films. As expected, the FDTD simulation confirms additional absorption in the gold film for wavelengths above the surface plasmon resonance wavelength, in qualitative agreement with measured transmission spectra. The cracks in the gold film exhibit intense wavelength-dependent hotspots with a typical local field intensity enhancement up to ≈30x in the absorbing regime. While unsuitable as semi-transparent coatings, such broad bandwidth nanoplasmonic systems are of importance in other fields.[35]

It should be noted that below the continuity threshold, the structure of the as-deposited film is unstable and XRR reveals decreasing film density (increased effective thickness) with time over a period of 6 months, with associated changes in transmission spectra (increased plasmonic absorption). On the other hand, for deposition thickness of 5.4 nm and above, no difference in thickness or density was measured in XRR over a period of 6 months (Figure 8). In the present work, we did not study the thermal stability of our films at elevated temperatures. However, a previous study of gold films deposited on optical polymers[30] confirmed that films below the continuity threshold exhibited a strong *increase* in sheet resistivity with time, while films above the continuity threshold remained continuous and showed a slight *decrease* in resistivity with time, also upon annealing at temperatures up to 150°C.

In summary, the above findings illustrate the importance of taking into account the transitional regime between Au and the surrounding materials, especially when optical properties of metal films in the nanometer-thick regime are considered, both for practical applications and when theoretical predictions are verified experimentally.

## 4. Conclusion

We have demonstrated that continuous ultra-thin gold films down to a mass-equivalent thickness of 5.4 nm can be realized directly on fused silica substrates, following surface activation with (3-mercaptopropyl)trimethoxysilane, using fast evaporation or high-energy irradiation prior to or during gold deposition. We show that optical transmission of such films is significantly lower than expected for an ideal film of the same mass-equivalent thickness and that optical transmission cannot, in this case, be accounted for by a simple one-layer model. This decreased transmission of continuous ultra-thin films can be accounted for, however, in a conventional transfer-matrix calculation by assuming "diffuse interfaces" on each side of the gold film. Presence of such interface layers is explicitly confirmed by X-ray reflectivity measurements. For thinner films in the coalescence regime, transmission is additionally affected by broad bandwidth plasmonic absorption, associated with geometrical features. The observed reduction of optical transmission is particularly important in the context of using ultra-thin gold films as transparent electrodes, where maximizing optical transparency is crucial. Furthermore, our results show that in order to truly separate the effects of structural (e.g. atomic-scale roughness) and physical (e.g. non-locality and quantum size effects) contributions to the dielectric response in ultra-thin films, it is necessary to fabricate atomically smooth gold films in this thickness regime, epitaxially grown onto atomically flat substrates. To our knowledge, such experiments have not yet been performed.

**Acknowledgements**

The authors thank Dr. Joachim F. Woitok from PANalytical, Netherlands for his advice on rocking curve fitting and Prof. Krishna Muralidharan for supplying DFT results. The project was supported by the Icelandic Research Fund, grants no. 110004021, 70021021 and 100019011, and by the UK Engineering and Physical Science Research Council's Active Plasmonics programme.

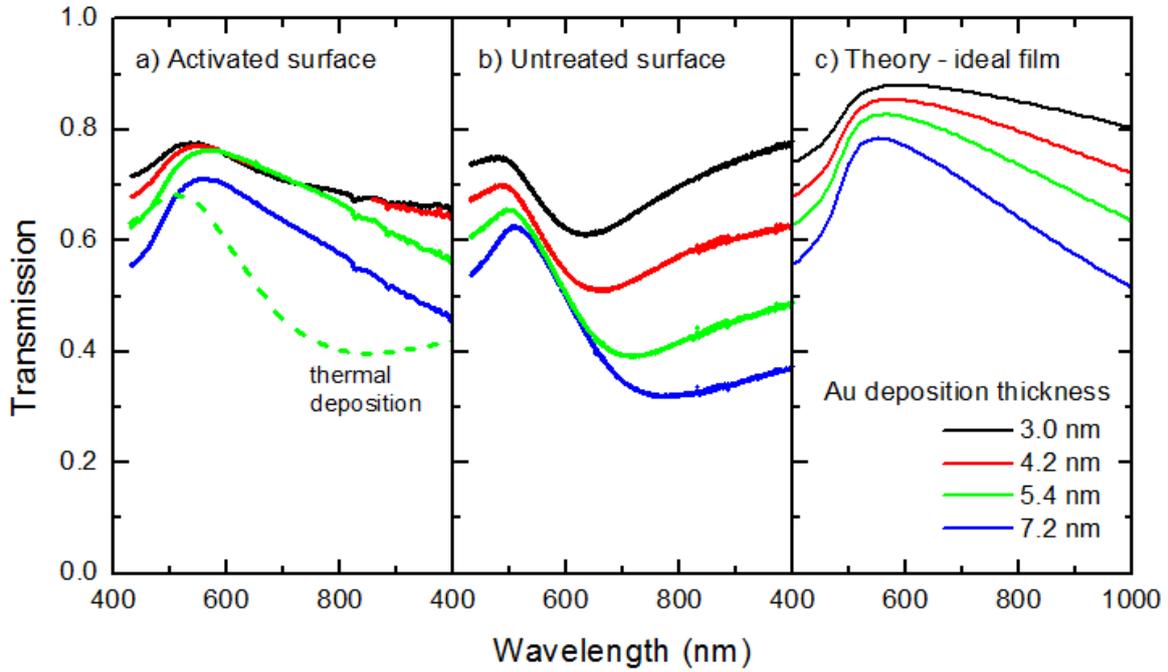

**Figure 1.** Optical transmission spectra through ultra-thin gold films of the indicated mass-equivalent thickness, simultaneously deposited on (a) (3-mercaptopropyl)trimethoxysilane activated fused silica and (b) untreated fused silica. (c) Theoretical transmission curves of ideal films calculated by the transfer-matrix method, using the bulk dielectric function of gold.

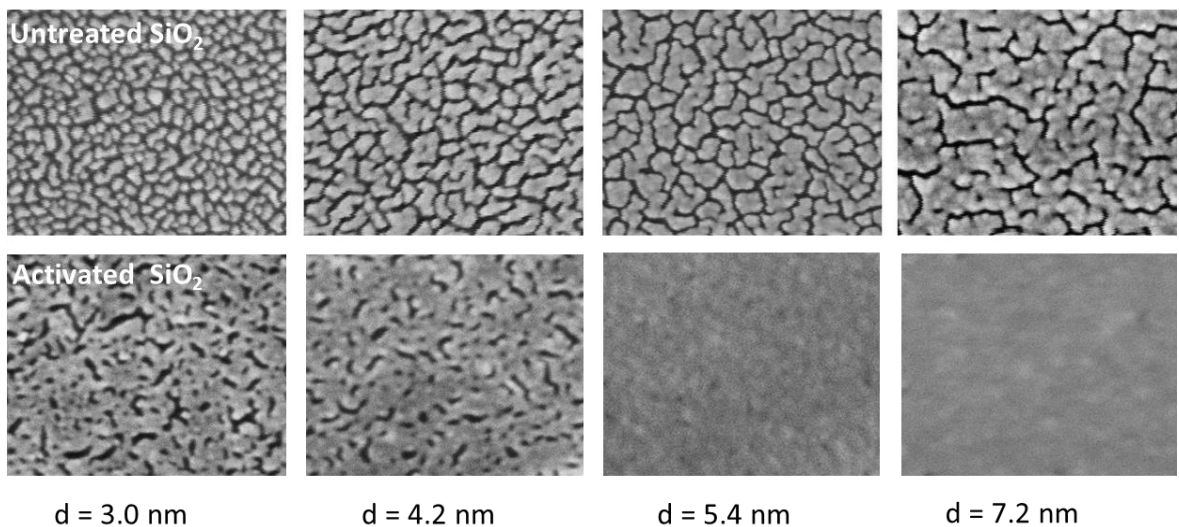

**Figure 2**. Scanning electron microscope images of gold films deposited on untreated SiO$_2$ (top row) and on fused silica substrates pre-treated with (3-mercaptopropyl)trimethoxysilane. On untreated glass, gold islands remain disconnected up to a deposition thickness of 5-7 nm, while coalescence occurs at a deposition thickness of 2-3 nm for activated glass. All images are 300 nm across.

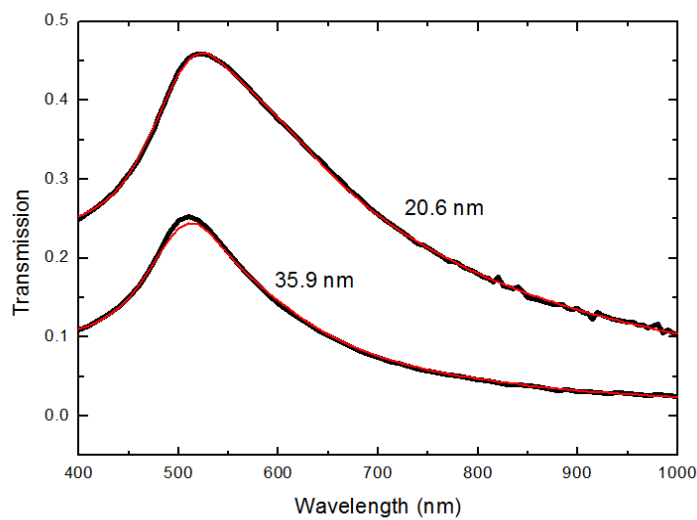

**Figure 3**. Optical transmission through fused silica coated with a layer of gold with the indicated thickness. Thick black curves represent measured data; thinner red curves are the results of transfer-matrix modelling assuming a single layer of gold, showing excellent agreement over the full measurement range.

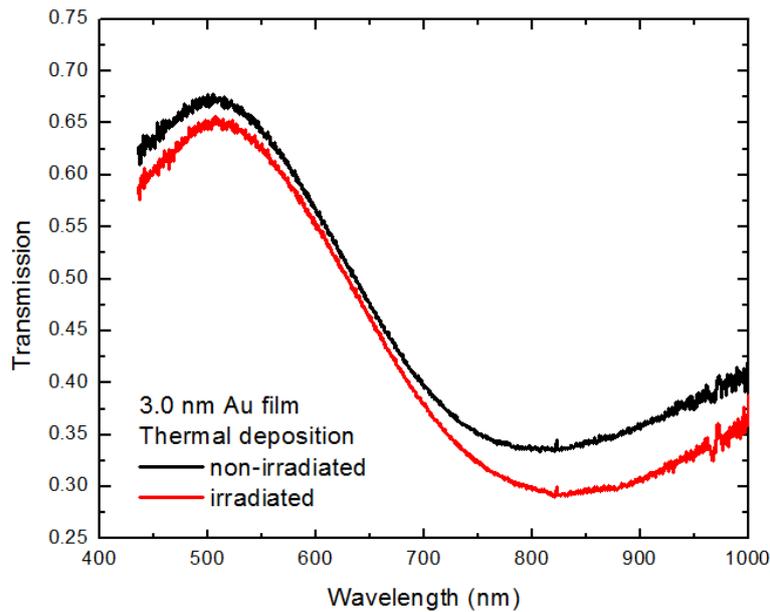

**Figure 4**. Transmission measured through gold films (island films) deposited with thermal evaporation onto fused silica substrates coated with (3-mercaptopropyl)trimethoxysilane with and without prior irradiation by UV light. Half of a thiol-activated substrate was covered by Al foil and exposed to UV light right before deposition. Subseqently, the Al foil was removed and the exposed and non-exposed areas of the slide were coated simultaneously. Although the effect is much less pronounced than that observed when using e-beam evaporation, where higher-energy electromagnetic radiation and secondary electron irradiation are involved, this simple measurement confirms that the attachment of gold to the prepared surface is visibly affected by irradiation prior to deposition.

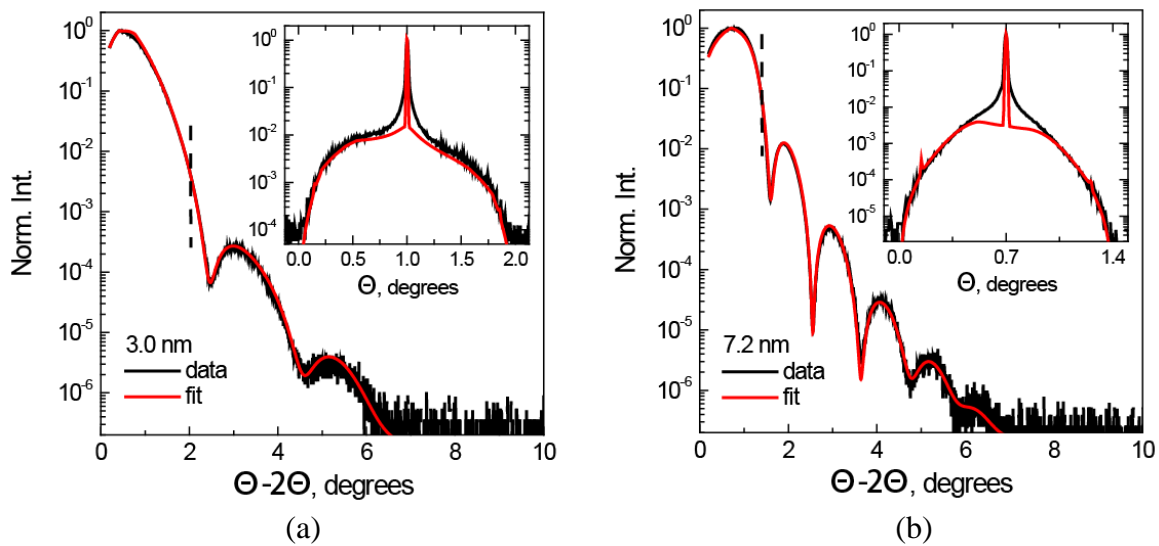

(a)   (b)

**Figure 5.** X-ray reflectivity spectra and the corresponding fitting curves for films of (a) 3.0 nm and (b) 7.2 nm mass-equivalent thickness, deposited on fused silica activated with (3-mercaptopropyl)trimethoxysilane. Excellent fits are obtained with a two-layer model for the gold film, as described in the text. The insets show rocking curves at the angular positions indicated by the dashed lines on the XRR curves.

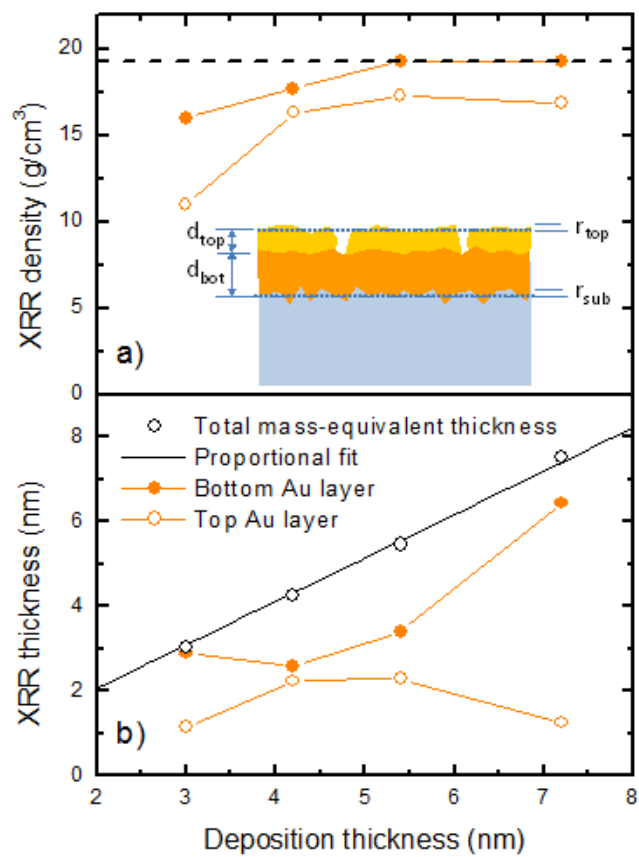

**Figure 3.** Fitting parameters derived from X-ray reflectivity spectra of gold films on activated fused silica, showing (a) density and (b) thickness of the two layers included in the model, as shown in the inset. The black open circles in the bottom panel indicate the total mass-equivalent thickness calculated from the XRR fitting parameters and the solid black line is a proportional fit to the data, showing excellent agreement with the QCM-measured deposition thickness.

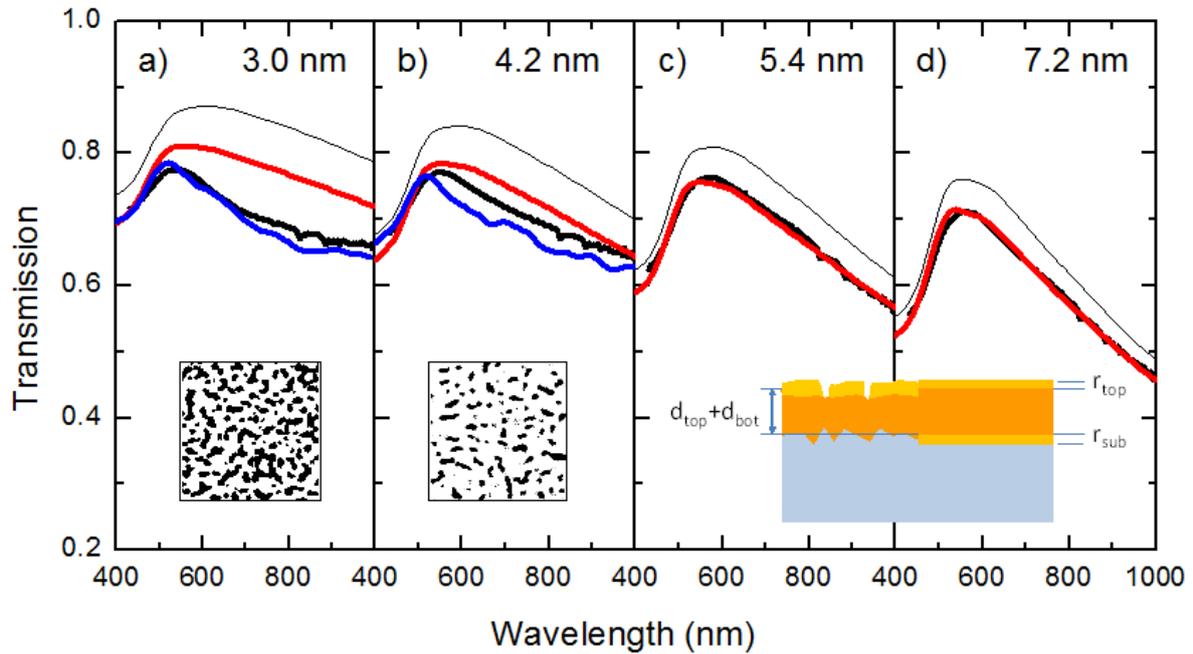

**Figure 7.** Optical transmission spectra of ultra-thin gold films. In the continuity regime (panels c and d), the measured data (thick black lines) agrees well with a three-layer model (red lines) that includes diffuse top and bottom interface layers, as indicated in the right half of the schematic figure. In the coalescence regime (panels a and b), additional plasmonic absorption is reproduced in FDTD simulations (blue lines), using a film structure derived from filtered SEM images of the corresponding films (insets show example images representing 200 x 200 $nm^2$ areas of the corresponding gold films (white=gold, black=voids)). Calculated transmission through ideal films of the same mass-equivalent thickness is shown for comparison (thin black lines).

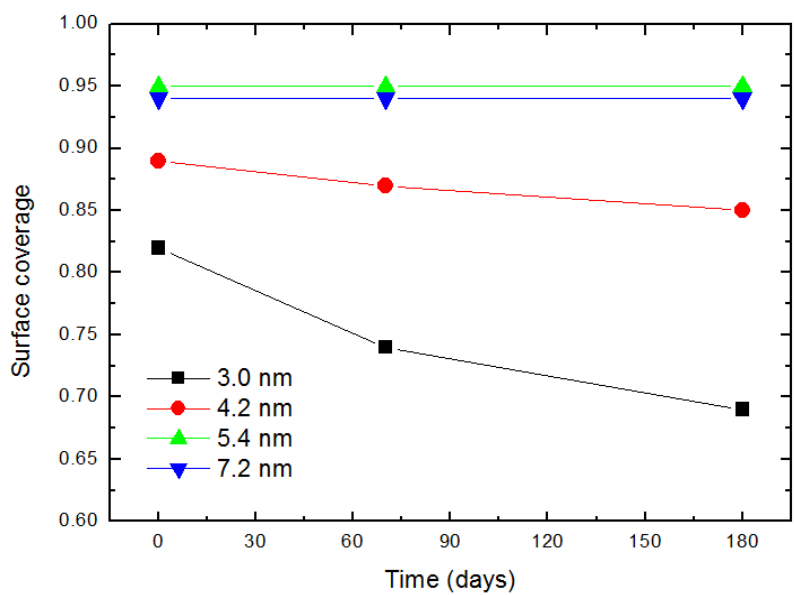

**Figure 8.** Development of the XRR-determined overall surface coverage, $d_{QCM}/(d_{top} + d_{bot})$, for films deposited onto (3-mercaptopropyl)trimethoxysilane-activated glass. Samples were stored under ambient conditions between measurements. For samples below the continuity threshold, the surface coverage clearly decreases with time, indicative of surface diffusion leading to increased island formation. Continuous films, however, remained completely stable during the investigated 6-month interval.